# Substrate-free THz focal plane metamaterial array with high absorption ratio

Zhigang Li,[1,2] Yi Ou[2], Xubiao Peng[1,4], Jianyu Fu[2], Wen Ou[2], Wenjing Jiang and Qing Zhao[1,3,5]

[1]*Center for Quantum Technology Research, School of Physics, Beijing Institute of Technology, Beijing, 100081, China*
[2]*Institute of Microelectronics of The Chines Academy of Sciences, Beijing, 100029, China*
[3]*Beijing Academy of Quantum Information Sciences, Beijing 100193, China*
[4]*xubiaopeng@bit.edu.cn*
[5]*qzhaoyuping@bit.edu.cn*

**Abstract:** Microelectromechanical system (MEMS) focal plane array (FPA) with optical readout offers exciting opportunities for real-time terahertz (THz) imaging. However, conventional FPA suffers from a low THz absorption ratio, which further decreases the performance of THz imaging. Here, we present a simple and scalable approach for the realization of THz focal plane metamaterial array with a relatively high absorption ratio. The key idea is to combine the advantages of substrate-free structures with metamaterial. A 100 × 100 THz FPA with a 150 × 150 μm pixel is designed, fabricated, and characterized. The dependence of the THz absorption ratio on the thickness of $SiN_x$ dielectric substrate film is investigated. The fabricated FPA exhibits a 90.6% resonant absorption at 1.36 THz, agreeing considerably with the theoretical simulation results. Our results imply that such a substrate-free THz focal plane metamaterial array enables the realization of THz imaging.

## 1. Introduction

Technology development of components and instrumentation for the largely unexplored terahertz (THz) spectral region ($10^{11}$–$10^{13}$ Hz) has proceeded rapidly in the last fifteen years [1–3]. It has attracted considerable attention because of its unique advantages, such as the strong ability of THz waves to penetrate nonpolarized objects and the fact that this region of the spectrum contains characteristic absorption bands of many materials and molecules. The THz poses less harmful effects on the probed persons or objects except the associated heating effects [4]. This feature gives THz a significant advantage over the traditional probing method, such as using x-rays. In addition, the THz detectors are widely applied in the detection of concealed objects as well as in non-invasive medical imaging techniques [5,6].

Metamaterial (MM) electromagnetic wave absorbers have shown great applications across numerous electromagnetic spectra and exhibit the ability to tailor the frequency-dependent emissivity and absorptivity [7,8]. Bi-material based substrate-free THz focal plane array (FPA) uses the MM absorbers for THz energy accumulation, and then transfer the absorbed energy into the small deformation of bi-material cantilevers [9,10].

The Bi-material based THz FPA can be used in THz imaging systems [11] when incorporated with an optical readout system, where the serpentine bi-material cantilevers are used as sensitive components. Fabio Alves et al. focused on the sensitive system of $SiO_2$/Al bimaterial cantilever. Their metamaterial structure is also based on the $SiO_2$/Al material system. At present, the absorption center frequency of $SiO_2$/Al metamaterials is generally higher than 3 THz. Similar studies with central absorption frequency below 3 THz have not been reported. For the super absorption frequency below 3 THz, we use the sensing system of

SiN$_x$/Au bimaterial cantilever and its metamaterial absorption structure. In this study, we have designed and developed a substrate-free THz focal plane metamaterial array with SiN$_x$/Au based bi-material cantilevers for real-time THz imaging.

## 2. Principle and Device Structure

A series of researches on THz metamaterial have been reported previously [12–14]. Those studies mainly focused on electromagnetic response analysis. In this work, a practical MEMS THz FPA with a metamaterial absorber was introduced, and the absorption of the detector was particularly investigated. It can be applied in THz imaging with an optical readout. The principle of the system is as follows: when the incident THz radiation is absorbed, the deflection is induced by the bi-material effect and can be further detected by the optical readout system, and finally, it is translated into a visible image.

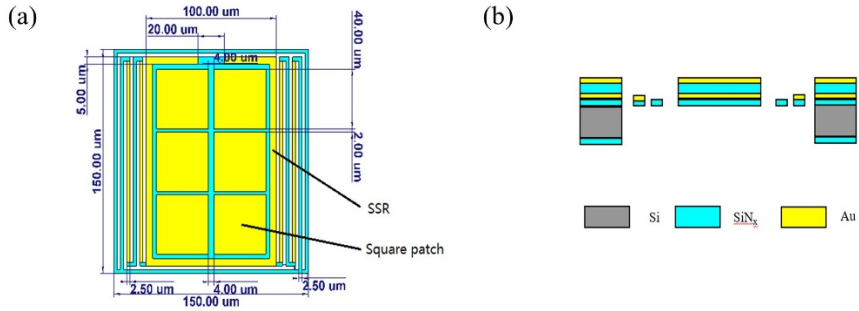

Fig. 1. (a) Top view of the pixel. (b) Cross-sectional view of the pixel.

The scheme of a pixel in the THz FPA is shown in Fig. 1. In the cross-section view of the pixel, from bottom to top, it consists of four layers: the lowest is used as the structure layer, including cantilevers and plates; the next one is a metal layer acting as a reflecting mirror for the optical readout system; the third layer is the absorption layer for THz radiation; the top layer is used to pattern the metamaterial structure.

The incident THz radiation from the target source is illuminated on the front of the FPAs. The absorbed radiation causes an increase in temperature, which results in the deformation of the micro-cantilever pixels in the array. Conversely, an LED is illuminated on the backside of the FPA. Then the light reflected from the FPA creates a thermal image and is recorded by a charge-coupled device (CCD) camera. The detailed operating principle was described in our previous work [9].

## 3. Design and Simulation

To characterize the electromagnetic response as a function of orientation, terahertz time-domain spectroscopy (THz-TDS) [15] was performed at a frequency range of 0.1–3 THz. All the analyses and discussions below are within this frequency range.

The low-stress silicon nitrogen film deposited by low-pressure chemical vapor deposition (LPCVD) was chosen as the structure layer of the FPA, due to its simple fabrication process and relatively high heat capacity. Afterward, plasma-enhanced chemical vapor deposition (PECVD) SiN$_x$ film was used as the absorption layer, and the reflecting mirror and metamaterial structure were made of gold. Hence, the bi-material cantilevers comprised the LPCVD SiN$_x$ film and the gold film.

To improve the sensitivity of the detector, two limitations need to be considered. First, how to increase the absorption at expected THz frequency, and second, how to increase the temperature as much as possible. Ideally, a periodic array is an effective way to enhance the absorption, and the factors, such as the proper shape, thickness, and metallic properties, enable high absorption at a specific frequency. In practice, the resonant absorption can be affected by the dimension of a patch of the array and thickness of the dielectric film beneath the periodic array, which is a form of frequency selective surfaces (FSS). According to the equivalent circuit method, the resonant wavelength of FSS is related to the effective electric length of the FSS patch element by the following approximate formula:

$$L_e = \frac{\lambda}{2\sqrt{\varepsilon_r}} = \frac{c}{2f\sqrt{\varepsilon_r}} \tag{1}$$

where $f$ and $\lambda$ are the frequency and wavelength of the incident electromagnetic wave, respectively, and $\varepsilon_r$ and c are the dielectric constant of the substrate and velocity of light, respectively. The effective length of the FSS patch was calculated to be approximately 35 μm at the frequency of 1.3 THz.

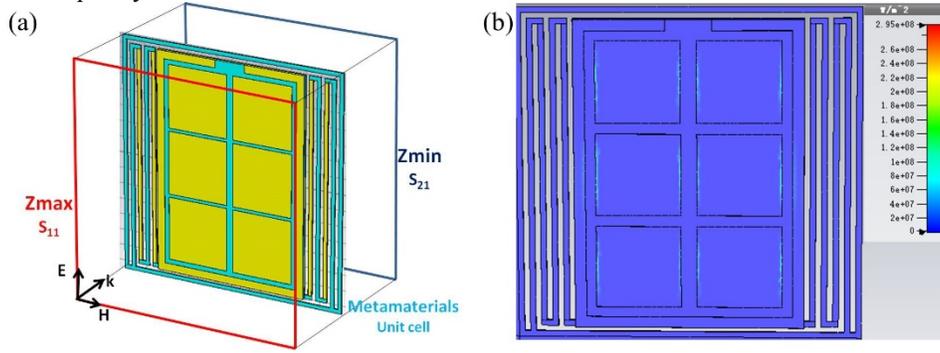

Fig. 2. (a) Metamaterial unit cell. (b) Power loss density distribution in the THz FPA.

To better understand the measured response, a CST microwave studio was used to simulate the electromagnetic response [16]. The structure is depicted in Fig. 1. The FSS, metamaterial, and unit cell module configuration, as shown in Fig. 2, was used. The metamaterial unit cell was sandwiched between air regions with $Z_{max}$ and $Z_{min}$ ports, as indicated in Fig. 2. To improve the accuracy of the model, actual dimensions were used, including thicknesses of the metallic layers. Material properties were taken from Table 1. A perpendicular plane wave was sent through the $Z_{max}$ port1, and the scattering parameters, $S_{11}$ and $S_{21}$, were extracted over the range of frequencies of interest. Reflectivity and transmissivity are given by $|S_{11}|^2$ and $|S_{21}|^2$, respectively. Absorptivity of the metamaterial is obtained directly by $1 - |S_{11}|^2 - |S_{21}|^2$. The power loss density in THz FPA is shown in Fig. 2(b). Simulations reveal that majority of the energy is dissipated as a dielectric loss at the edge of the square patch.

**Table 1. Properties of the constituent materials of the THz FPA**

| Material | Young's Modulus $E(GPa)$ | Expansion Coefficient $a(10^{-6}K^{-1})$ | Thermal Conductivity $g(Wm^{-1}K^{-1})$ | Heat Capacity $c(JKg^{-1}K^{-1})$ | Density $\rho(kgm^{-3})$ | Electric conductivity $\sigma(10^6 Sm^{-1})$ | Relative permittivity $\varepsilon$ |
|---|---|---|---|---|---|---|---|
| Si | 112 | 5.1 | 148 | 700 | 2330 | - | 11.9 |
| SiN$_x$ | 180 | 2.1 | 19 | 691 | 2400 | - | 7.5 |
| Au | 78 | 14 | 314 | 130 | 19320 | 45.61 | - |

The resonant frequency is mainly determined by the dimension of the patch array, i.e., the length of the patch side. In Fig. 3a, we present the simulation result on the relation between resonant frequency and patch side, where the curve is well fitted by a simple equation, $f = \frac{A}{L}$, according to equation (1), where $A = \frac{c}{2\sqrt{\varepsilon_r}} = 54.77\,\mu m/s$. In particular, when the patch side L= 40 μm, the response frequency $f$ = 1.36 THz, which is close to the frequency of the maximum spectral amplitude provided by the laser.

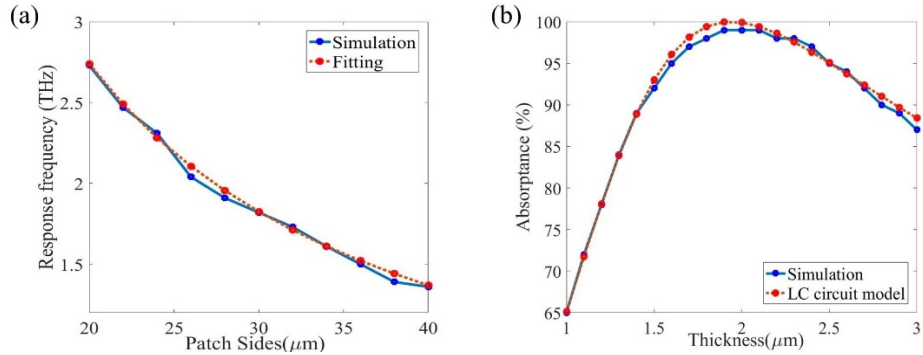

Fig. 3. (a) Relationship between patch sides and response frequency, where the blue line is the simulation result, and the red dashed line is the fitting curve. (b) Relationship between the thickness of the silicon nitrogen film and absorption.

Since the absorption can also be affected by the dielectric layer beneath the FSS, we also investigated the relationship between the $SiN_x$ film thickness and absorption at the expected frequency range obtained from simulations. The result is shown in Fig. 3b. It shows that the highest theoretical absorption level is 99.36% when the film thickness is 2 μm. We note that such a relation can be modeled by a simple equivalent LC circuit described in the literature [17], as indicated by the red-lined curve in Fig. 3b. The detailed parameters for the LC circuit model are described in Supplemental Material.

Next, with $SiN_x$ film thickness at 2 μm and patches side length at 40 μm, the effects of the split ring resonator (SSR) in FPA was investigated. The resonant frequency of the SSR was determined by the ring length and the split size. In this work since the SSR is much longer than the patch side, the resonant frequency of the SSR is much lower than that of the FSS. Hence, the SSR has barely any effect on the response frequency and the absorption of the pixel. In Fig. 4, we show that there is a slight difference in the absorptions of FPA with different SSR split sizes from the simulation results.

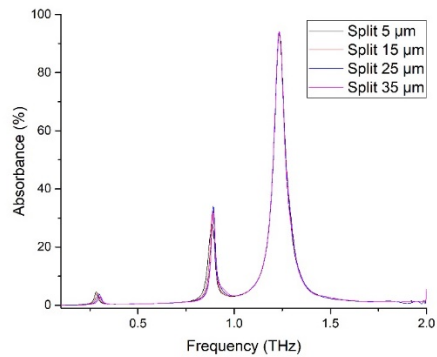

Fig. 4. Simulated absorption with deferent SSR split sizes.

Besides the metamaterial structure, another way to improve the sensitivity of the FPA is to use the CCD with a high quantization level or to amplify the pixel deflection caused by the bi-material effect. The detection sensitivity of the optical readout system was determined by the minimum detectable angle $\theta_{min}$ corresponding to one grey level of CCD. It can be expressed as [9]:

$$\theta_{min} = \frac{\Delta\theta}{\Delta I} = \frac{\lambda}{2LN} \qquad (2)$$

where $\lambda$ is the wavelength of the readout light, $L$ is the reflector length along the cantilever direction, $N$ is the quantization level of CCD, and $\Delta\theta$ and $\Delta I$ are the changes in the deformation angle and the grey level of CCD, respectively. Therefore, the reflector length along the leg is required to be as long as possible to minimize $\theta_{min}$. Contrarily, a larger deformation angle can be achieved by increasing the number of folded cantilevers. We note that both methods above are limited by fabrication, device volume, and image resolution.

## 4. Fabrication and Measurement

All the samples were prepared with conventional micro/nano-fabrication processes, as shown in Fig. 5. The fabrication process flow is as follows: a. LPCVD SiN$_x$ layer with a thickness of 5000 Å was generated on double sides of the wafer; b. 1000 Å gold layer was evaporated on the front side; c. 1.8 μm PECVD SiN$_x$ layer was deposited on top of the gold layer; d. lithography and reactive ion etching (RIE) were performed to open the back etching window; e. Front lithography and RIE were performed to pattern the absorption layer below the metamaterial structure; f. Cantilevers etching; g. Cantilevers and metamaterial structure patterning by gold evaporation; h. Pixel release by backside silicon wet etching.

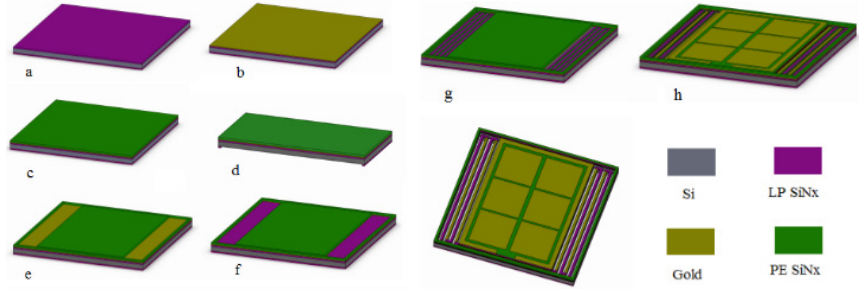

Fig. 5. Fabrication flowchart of the MEMS THz FPA.

An example of the fabricated free-standing THz FPA is shown in Fig. 6. The obtained FPA is 1.5 × 1.5 cm and holds an array of 100 × 100 pixels.

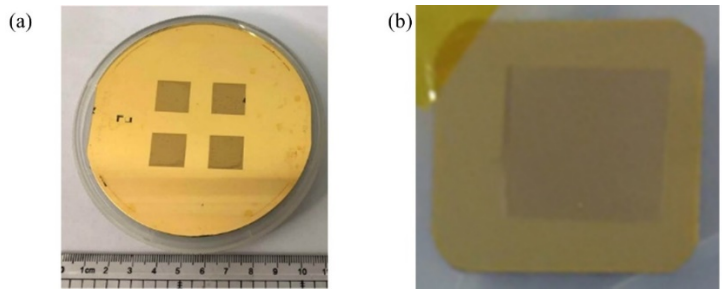

Fig. 6. (a) Photograph of four fabricated THz FPA on a 4-inch silicon wafer. (b) Fabricated and diced THz FPA chip of 15 × 15 mm.

In Fig. 7, we show the three-dimensional shape of pixels observed by 3D measuring laser microscope and the side view of the reflector, respectively. Using the low-stress silicon nitride film deposited by LPCVD at 900 °C, the smooth reflector was obtained, which contributed to the optical readout.

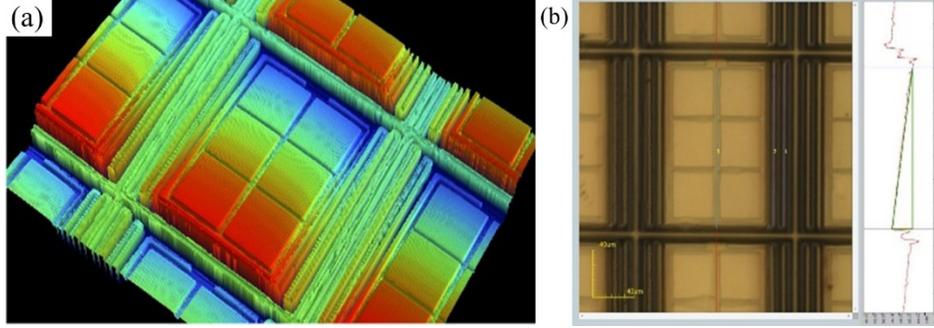

Fig. 7. (a) Three-dimensional shape of pixels. (b) Measured flatness of the reflectors.

Thermal sensitivity was measured using a hot plate and three-dimensional (3D) measuring laser microscope. As shown in Fig. 8, the deformation angle of the reflector is the result of angles from both the 1st and 2nd cantilevers, proving the amplifying design works. The thermal sensitivity was calculated to be 0.075 °/°C according to a linear fitting on the deformation angles of the reflector in Fig. 8a, which is greater than that of the infrared (IR) FPA with 0.051 °/°C as we reported previously [9]. A consideration of the conservation of energy allows us to describe the increase in temperature of the metamaterial as [18]:

$$\Delta T = \frac{\alpha_m P_m}{\sqrt{G_T^2 + \omega^2 C^2}} \tag{3}$$

where $\alpha_m$ is the emittance/absorbance of the metasurface, $P_m$ is the incident THz power modulated at a specific frequency $\omega$, $C$ is the heat capacity of the metasurface, and $G_T$ is the thermal conductance of the total power loss, which consists of convection, conduction, and radiation. With the method described in [9], we obtained the thermal sensitivity, $4.96 \times 10^5$ K/W, at the same modulated frequency, 0.25 Hz, which is approximately 20 times more than KEBIN FAN's reported [18].

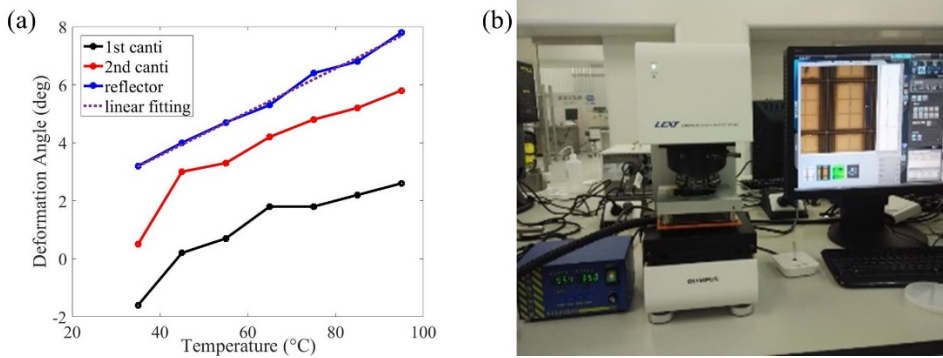

Fig. 8. (a) Measured thermal sensitivity of THz FPA. The purple dashed line is a fitting of the deformation angle of the reflector using linear function $y = 0.075x + 0.56$. (b) Measured equipment.

Finally, FPA absorption for THz radiation was measured using a THz time-domain spectroscopy system, which was reported by Xiaojun Wu [19]. The incident angle of the THz beam was fixed at 20° and 30°. In Fig. 9, an absorption peak of 90.6% at 1.36 THz was observed. From the comparison between the measurement and simulation results, a good agreement can be observed between them.

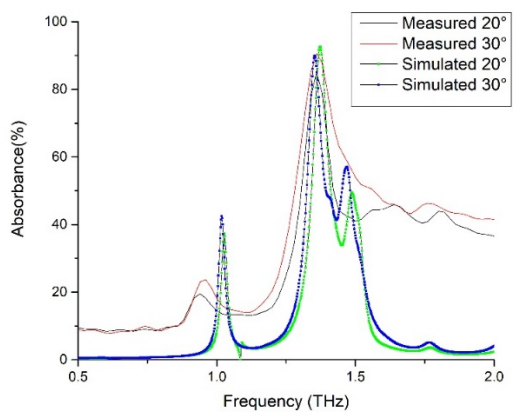

Fig. 9. Measured absorption against frequency.

## 5. Conclusion

By investigating the relationship between the absorption and thickness of the $SiN_x$ dielectric film, we selected the optimal thickness for THz absorption. Afterward, a 100 × 100 THz FPA with a pixel of 150 × 150 μm and a fill factor of 66.7% was successfully designed. Through direct measurement, we find that it has higher thermal sensitivity (up to 0.075°/°C) than IR FPA. Furthermore, its absorption rate is up to 90.6% at 1.36 THz, which is consistent with the simulation result. Due to the above observed high thermal sensitivity and absorption ratio, we believe that this kind of FPA has a powerful potential application in THz imaging.

## Funding

National Natural Science Foundation of China (61874137).

## Acknowledgments

Zhigang Li and Yi Ou contributed equally to this work. The authors would like to thank Prof. Li Wang for the THz-TDS measurement.

## Disclosures

The authors declare no conflicts of interest.

See Supplement 1 for supporting content.